\input harvmac.tex

\vskip 1.5in
\Title{\vbox{\baselineskip12pt
\hbox to \hsize{\hfill}
\hbox to \hsize{\hfill }}}
{\vbox{
	\centerline{\hbox{String Amplitudes and
		}}\vskip 5pt
        \centerline{\hbox{ Frame-Like Formalism for Higher Spins
		}} } }
\centerline{Seungjin Lee$^{2}$}
\centerline{and}
\centerline{Dimitri Polyakov$^{1}$\footnote{$^\dagger$}
{polyakov@sogang.ac.kr, dimitri.polyakov@wits.ac.za, 
twistorstring@gmail.com;
on leave of absence from National Institute
for Theoretical Physics and School of Physics, University of
the Witwatersrand, WITS 2050 Johannesburg, South Africa}}
\medskip
\centerline{\it Center for Quantum Space-Time (CQUeST)$^{1}$}
\centerline{\it and Department of Physics$^{2}$}
\centerline{\it Sogang University}
\centerline{\it Seoul 121-742, Korea}

\vskip .3in

\centerline {\bf Abstract}

We analyze open string vertex operators describing
connection gauge fields for spin 3
in Vasiliev's frame-like formalism
and perform their extended BRST analysis.
Gauge symmetry transformations,
generalized zero torsion constraints relating
extra fields to the dynamical frame-like field and 
relation between dynamical frame-like field
and fully symmetric Fronsdal's field for spin 3
are all realized in terms of BRST constraints
on these vertex operators in string theory.
Using the construction, we analyze the 3-point 
correlator for spin 3 field and calculate
 Chern-Simons type cubic interactions
described by  $3$-derivative
Berends-Burgers-Van Dam (BBD) type vertex
in the frame-like formalism.

\Date{March 2012}

\vfill\eject

\lref\fvf{E.S. Fradkin, M.A. Vasiliev, Nucl. Phys. B 291, 141 (1987)}
\lref\fvs{E.S. Fradkin, M.A. Vasiliev, Phys. Lett. B 189 (1987) 89}
\lref\mmswf{S.W. MacDowell, F. Mansouri, Phys. Rev.Lett. 38 (1977) 739}
\lref\mmsws{K. S. Stelle and P. C. West, Phys. Rev. D 21 (1980) 1466}
\lref\mmswt{C.Preitschopf and M.A.Vasiliev, hep-th/9805127}
\lref\vmaf{M. A. Vasiliev, Sov. J. Nucl. Phys. 32 (1980) 439,
Yad. Fiz. 32 (1980) 855}
\lref\vmas{V. E. Lopatin and M. A. Vasiliev, Mod. Phys. Lett. A 3 (1988) 257}
\lref\vmat{E.S. Fradkin and M.A. Vasiliev, Mod. Phys. Lett. A 3 (1988) 2983}
\lref\vmafth{M. A. Vasiliev, Nucl. Phys. B 616 (2001) 106 }
\lref\bianchi{ M. Bianchi, V. Didenko, arXiv:hep-th/0502220}
\lref\ruehl{R. Manvelyan and W. Ruehl, hep-th/0502123}
\lref\bonellio{G. Bonelli, JHEP 0411 (2004) 059}
\lref\hsaone{M. A. Vasiliev, Fortsch. Phys. 36 (1988) 33}
\lref\hsatwo{E.S. Fradkin and M.A. Vasiliev, Mod. Phys. Lett. A 3 (1988) 2983}
\lref\hsathree{S. E. Konstein and M. A. Vasiliev, Nucl. Phys. B 331 (1990) 475}
\lref\hsafour{M.P. Blencowe, Class. Quantum Grav. 6, 443 (1989)}
\lref\hsafive{E. Bergshoeff, M. Blencowe and K. Stelle, 
Comm. Math. Phys. 128 (1990) 213}
\lref\hsasix{E. Sezgin and P. Sundell, Nucl. Phys. B 634 (2002) 120 }
\lref\hsaseven{M. A. Vasiliev, Phys. Rev. D 66 (2002) 066006 }
\lref\soojongf{M. Henneaux, S.-J. Rey, JHEP 1012:007,2010}
\lref\henneaux{J. D. Brown and M. Henneaux, Commun. Math. Phys. 104, 207 (1986)}
\lref\sagnottia{A. Sagnotti, E. Sezgin, P. Sundell, hep-th/0501156}
\lref\sorokin{D. Sorokin, AIP Conf. Proc. 767, 172 (2005)}
\lref\fronsdal{C. Fronsdal, Phys. Rev. D18 (1978) 3624}
\lref\coleman{ S. Coleman, J. Mandula, Phys. Rev. 159 (1967) 1251}
\lref\haag{R. Haag, J. Lopuszanski, M. Sohnius, Nucl. Phys B88 (1975)
257}
\lref\weinberg{ S. Weinberg, Phys. Rev. 133(1964) B1049}
\lref\tseytbuch{E. Buchbinder, A. Tseytlin, JHEP 1008:057,2010}
\lref\fradkin{E. Fradkin, M. Vasiliev, Phys. Lett. B189 (1987) 89}
\lref\skvortsov{E. Skvortsov, M. Vasiliev, Nucl.Phys.B756:117-147 (2006)}
\lref\skvortsovb{E. Skvortsov, J.Phys.A42:385401 (2009)}
\lref\mva{M. Vasiliev, Phys. Lett. B243 (1990) 378}
\lref\mvb{M. Vasiliev, Int. J. Mod. Phys. D5
(1996) 763}
\lref\mvc{M. Vasiliev, Phys. Lett. B567 (2003) 139}
\lref\brink{A. Bengtsson, I. Bengtsson, L. Brink, Nucl. Phys. B227
 (1983) 31}
\lref\deser{S. Deser, Z. Yang, Class. Quant. Grav 7 (1990) 1491}
\lref\bengt{ A. Bengtsson, I. Bengtsson, N. Linden,
Class. Quant. Grav. 4 (1987) 1333}
\lref\boulanger{ X. Bekaert, N. Boulanger, S. Cnockaert,
J. Math. Phys 46 (2005) 012303}
\lref\bbd{F. Berends, G. Burgers, H. Van Dam ,Nucl.Phys. B260 (1985) 295}
\lref\metsaev{ R. Metsaev, arXiv:0712.3526}
\lref\siegel{ W. Siegel, B. Zwiebach, Nucl. Phys. B282 (1987) 125}
\lref\siegelb{W. Siegel, Nucl. Phys. B 263 (1986) 93}
\lref\nicolai{ A. Neveu, H. Nicolai, P. West, Nucl. Phys. B264 (1986) 573}
\lref\damour{T. Damour, S. Deser, Ann. Poincare Phys. Theor. 47 (1987) 277}
\lref\sagnottib{D. Francia, A. Sagnotti, Phys. Lett. B53 (2002) 303}
\lref\sagnottic{D. Francia, A. Sagnotti, Class. Quant. Grav.
 20 (2003) S473}
\lref\sagnottid{D. Francia, J. Mourad, A. Sagnotti, Nucl. Phys. B773
(2007) 203}
\lref\labastidaa{ J. Labastida, Nucl. Phys. B322 (1989)}
\lref\labastidab{ J. Labastida, Phys. Rev. Lett. 58 (1987) 632}
\lref\mvd{L. Brink, R.Metsaev, M. Vasiliev, Nucl. Phys. B 586 (2000) 183}
\lref\klebanov{ I. Klebanov, A. M. Polyakov,
Phys.Lett.B550 (2002) 213-219}
\lref\mve{
X. Bekaert, S. Cnockaert, C. Iazeolla,
M.A. Vasiliev,  IHES-P-04-47, ULB-TH-04-26, ROM2F-04-29, 
FIAN-TD-17-04, Sep 2005 86pp.}
\lref\sagnottie{A. Campoleoni, D. Francia, J. Mourad, A.
 Sagnotti, Nucl. Phys. B815 (2009) 289-367}
\lref\sagnottif{
A. Campoleoni, D. Francia, J. Mourad, A.
 Sagnotti, arXiv:0904.4447}
\lref\sagnottig{D. Francia, A. Sagnotti, J.Phys.Conf.Ser.33:57 (2006)}
\lref\selfa{D. Polyakov, Int.J.Mod.Phys.A20:4001-4020,2005}
\lref\selfb{ D. Polyakov, arXiv:0905.4858}
\lref\selfc{D. Polyakov, arXiv:0906.3663, Int.J.Mod.Phys.A24:6177-6195 (2009)}
\lref\selfd{D. Polyakov, Phys.Rev.D65:084041 (2002)}
\lref\spinself{D. Polyakov, Phys.Rev.D82:066005,2010}
\lref\spinselff{D. Polyakov,Phys.Rev.D83:046005,2011}
\lref\mirian{A. Fotopoulos, M. Tsulaia, Phys.Rev.D76:025014,2007}
\lref\extraa{I. Buchbinder, V. Krykhtin,  arXiv:0707.2181}
\lref\extrab{I. Buchbinder, V. Krykhtin, Phys.Lett.B656:253-264,2007}
\lref\extrac{X. Bekaert, I. Buchbinder, A. Pashnev, M. Tsulaia,
Class.Quant.Grav. 21 (2004) S1457-1464}
\lref \extrad{I. Buchbinder, A. Pashnev, M. Tsulaia,
arXiv:hep-th/0109067}
\lref\extraf{I. Buchbinder, A. Pashnev, M. Tsulaia, 
Phys.Lett.B523:338-346,2001}
\lref\extrag{I. Buchbinder, E. Fradkin, S. Lyakhovich, V. Pershin,
Phys.Lett. B304 (1993) 239-248}
\lref\extrah{I. Buchbinder, A. Fotopoulos, A. Petkou, 
 Phys.Rev.D74:105018,2006}
\lref\bonellia{G. Bonelli, Nucl.Phys.B {669} (2003) 159}
\lref\bonellib{G. Bonelli, JHEP 0311 (2003) 028}
\lref\ouva{C. Aulakh, I. Koh, S. Ouvry, Phys. Lett. 173B (1986) 284}
\lref\ouvab{S. Ouvry, J. Stern, Phys. Lett.  177B (1986) 335}
\lref\ouvac{I. Koh, S. Ouvry, Phys. Lett.  179B (1986) 115 }
\lref\hsself{D.Polyakov, arXiv:1005.5512}
\lref\sundborg{ B. Sundborg, ucl.Phys.Proc.Suppl. 102 (2001)}
\lref\sezgin{E. Sezgin and P. Sundell,
Nucl.Phys.B644:303- 370,2002}
\lref\morales{M. Bianchi,
J.F. Morales and H. Samtleben, JHEP 0307 (2003) 062}
\lref\giombif{S. Giombi, Xi Yin, arXiv:0912.5105}
\lref\giombis{S. Giombi, Xi Yin, arXiv:1004.3736}
\lref\bekaert{X. Bekaert, N. Boulanger, P. Sundell, arXiv:1007.0435}
\lref\taronna{A. Sagnotti, M. Taronna, arXiv:1006.5242, 
Nucl.Phys.B842:299-361,2011}
\lref\zinoviev{Yu. Zinoviev, arXiv:1007.0158}
\lref\fotopoulos{A. Fotopoulos, M. Tsulaia, arXiv:1007.0747}
\lref\fotopouloss{A. Fotopoulos, M. Tsulaia, arXiv:1009.0727}
\lref\taronnao{M. Taronna, arXiv:1005.3061}
\lref\taronnas{A. Sagnotti, M. Taronna, arXiv:1006.5242 ,
Nucl.Phys.B842:299-361,2011}
\lref\campo{A.Campoleoni,S. Fredenhagen,S. Pfenninger, S. Theisen,
arXiv:1008.4744, JHEP 1011 (2010) 007}
\lref\gaber{M. Gaberdiel, T. Hartman, arXiv:1101.2910, JHEP 1105 (2011) 031}
\lref\per{	
N. Boulanger,S. Leclercq, P. Sundell, JHEP 0808(2008) 056 }
\lref\mav{V. E. Lopatin and M. A. Vasiliev, Mod. Phys. Lett. A 3 (1988) 257}
\lref\zinov{Yu. Zinoviev, Nucl. Phys. B 808 (2009)}
\lref\sv{E.D. Skvortsov, M.A. Vasiliev,
Nucl. Phys.B 756 (2006)117}
\lref\mvasiliev{D.S. Ponomarev, M.A. Vasiliev, Nucl.Phys.B839:466-498,2010}
\lref\zhenya{E.D. Skvortsov, Yu.M. Zinoviev, arXiv:1007.4944}
\lref\perf{N. Boulanger, C. Iazeolla, P. Sundell, JHEP 0907 (2009) 013 }
\lref\pers{N. Boulanger, C. Iazeolla, P. Sundell, JHEP 0907 (2009) 014 }
\lref\selft{D. Polyakov,Phys.Rev.D82:066005,2010}
\lref\selftt{D. Polyakov, Int.J.Mod.Phys.A25:4623-4640,2010}
\lref\tseytlin{I. Klebanov, A Tseytlin, Nucl.Phys.B546:155-181,1999}
\lref\ruben{R. Manvelyan, K. Mkrtchyan, W. Ruehl, arXiv:1009.1054}
\lref\rubenf{R. Manvelyan, K. Mkrtchyan, W. Ruehl, Nucl.Phys.B836:204-221,2010}
\lref\robert{
R. De Mello Koch, A. Jevicki, K. Jin, J. A. P. Rodrigues, arXiv:1008.0633}
\lref\bek{X. Bekaert, S. Cnockaert, C. Iazeolla, M. A. Vasiliev,
hep-th/0503128}
\lref\vcubic{M. Vasiliev, arXiv:1108.5921}
\lref\sagnottinew{A. Sagnotti, arXiv:1112.4285}
\lref\yin{C.-M. Chang, X. Yin, arXiv:1106.2580 }
\lref\boulskv{ N. Boulanger, E. Skvortsov, arXiv:1107.5028,
JHEP 1109 (2011) 063}
\lref\boulskvz{N. Boulanger, E. Skvortsov,Yu. Zinoviev,
arXiv:1107.1872 , J.Phys.A A44 (2011) 415403
}
\lref\didenko{V. Didenko, Class.Quant.Grav. 29 (2012) 025009}
\lref\selfsigma{D. Polyakov, Phys.Rev. D84 (2011) 126004}
\lref\meinprogress{D. Polyakov, in progress}
\lref\sjprogress{S.-J. Rey and  D. Polyakov, in progress}

\centerline{\bf  1. Introduction}

Constructing consistent gauge theories
of interacting higher spin fields is a long-standing, fascinating
and difficult problem
(for an incomplete and very subjective list of references see

~{\fronsdal, \fvf, \fvs, \sagnottia, \sagnottib, \sagnottic,
\sagnottid, \sorokin, 
\mva, \mvb, \mvc, \mvasiliev, \deser, \bengt, \siegel, \siegelb,
\nicolai, \damour, \brink, \boulanger,
\labastidaa, \labastidab, 
\mvd, \mve, \mvasiliev, \mirian, \vcubic, \extrah,
\extrad, \bek, \bekaert, 
\perf, \pers, \taronna, \taronnao, 
\zinoviev,
\fotopoulos, \fotopouloss, \taronna,\zinoviev, \bekaert, \morales, \giombif, 
\giombis, \campo, \gaber, \boulskv, \boulskvz, 
\sezgin, \sundborg, \per, \zhenya, \robert, \hsafour, \hsafive, \hsasix,
\hsaseven, \soojongf, \yin, \henneaux, \campo, \gaber, \boulskv,
\boulskvz, \didenko, \bianchi})

 Despite significant progress
in describing the dynamics of higher spin field theories,
achieved over recent few decades, our understanding
of the general structure of the higher spin interactions
is still very far from complete. 
String theory appears to be a particularly efficient and natural
framework to construct and analyze consistent gauge-invariant
interactions of higher spins 
~{\sagnottia, \sagnottib, \sagnottic, \sagnottib, 
 \sagnottie, \sagnottif, \sagnottig, \taronnao, \taronnas 
\labastidaa, \labastidab,\metsaev,\ruben, \rubenf, 
\fotopoulos, \fotopouloss,
\spinself, \spinselff, \sagnottinew}

Within string theory, there are several approaches
to this problem. The first approach is based
on the observation that excitations with higher spins
appear naturally in the massive spectrum of open and closed
strings with the masses of the states on the leading Regge 
trajectory
given by $m\sim({s\over{\alpha^{\prime}}})^{1\over2}$,
so in the tensionless limit $\alpha^\prime\rightarrow\infty$
the corresponding operators technically become massless.
There are several difficulties within this approach,
e.g. it is generally not easy to combine
 the vertex operators so as to recover
the explicit set of the  Stueckelberg
symmetries  of the corresponding states.
The known examples of such operators typically
mix the excitations with different spin values ~{\sagnottinew}.
In addition, since the tensionless limit is opposite
 to the low energy one, field theoretic interpretation
of the correlation functions of these vertex operators
is not easy. 
This formalism is also hard to extend to the AdS 
case since the worldsheet correlators of string theory
in AdS backgrounds are difficult to analyze beyond
 the semiclassical limit.
Another string theoretic approach to higher spins, 
based on the formalism of
ghost cohomologies, is independent on the tension
arguments and in principle
 allows 
to circumvent some of the difficulties 
related to the tensionless
limit. This approach is based on new physical
(BRST invariant and nontrivial)vertex operators
that we analyzed in previous works (see e.g. ~{\spinself})
that are essentially coupled to the $\beta-\gamma$ system
of superconformal ghosts in RNS formalism. This ghost coupling
cannot be removed by picture-changing transformation
and can be classified in terms of ghost cohomologies ~{\spinself,
\spinselff}.
This class of vertex operators is ghost picture-dependent,
distinguishing them from standard operators such as
a photon or a graviton, that exist at any picture.
In open string sector, there is a subclass of these operators
corresponding to massless higher spin excitations.
BRST invariance conditions lead to Pauli-Fierz on-shell
conditions for higher spin fields in Fronsdal's metric-like 
formalism while BRST nontriviality constraints lead
to gauge transformations for these operators.
Their worldsheet amplitudes are thus gauge-invariant by
construction and describe polynomial interactions of
massless higher
spin fields in the low energy effective limit.
In our previous works we calculated some examples
of  such interations - cubic interaction
of $s=3-3-4$,
, the disc amplitude of spin 3 operators
with the graviton (reproducing the coupling of spin 3 to gravity
through linearized Weyl tensor) and the quartic interaction
of spin 3 and spin 1 gauge fields ~{\spinself, \spinselff}.
In practice, however, explicit calculations
involving these operators are 
in most cases
are complicated, as their explicit structure
is generally quite cumbersome. More significantly,
due to the picture dependence,
 in many physically important cases the options
to manipulate with the picture changing are limited and 
it is often hard 
to find the appropriate picture combination of the
the higher spin vertex operator satisfying the correct
ghost number balance in correlation functions to cancel
the background charges of the ghosts
(e.g. on the sphere all the appropriate correlators
must carry total $\phi$-ghost number $-2$, $\chi$-ghost
 number $+1$ and $b-c$ ghost number $+3$).
One important example when such complication appears
is the cubic interaction of spin $s=3$ corresponding
to cubic amplitude of spin 3 vertex operators in open string
theory. Straightforward calculation 
of this amplitude using vertex operators
for Fronsdal-type fields requires 4 picture changing transformations
which, given cumbersome structure
of the operators, makes the computations practically 
insurmountable.
In this paper we approach this problem by developing vertex 
operator formalism for auxiliary (extra)
 fields in Vasiliev's frame-like
approach. We construct vertex operators for connection
gauge fields  in this formalism. As in the Fronsdal's case
the on-shell conditions on the operators lead to standard
trace and symmetry constraints on the fiber indices
of the connection gauge fields, along with gauge fixing
conditions for diffeomorphism symmetries. Gauge transformations
of the connection fields lead, in turn, to shifting the
vertex operators by BRST exact terms that do not affect the correlators
that determine the structure of the interaction terms in the low-energy 
limit. The generalized zero torsion constraints follow from
ghost cohomology conditions on the vertex operators that will
be derived in the next section.

The rest of the paper is organized as follows.
In the Section 2 we review the basic ideas of the frame-like
 description of higher spin fields and construct vertex operators
for the dynamical and auxiliary connection gauge fields.
In the Section 3 we analyze the 3-point correlation function
of these operators for spin 3, limiting ourselves to terms
with 3 derivatives. The result is given by the 
Berends-Burgers-Van Dam type 3-derivative vertex in a certain gauge,
modulo total derivative terms.In the Discussion section we
outline generalizations of the developed formalism for AdS case 
and discuss the relation between the vertex operators,
constructed in this paper and generators of higher spin algebra
in AdS.

\centerline{\bf 2. Frame-like Formalism and Vertex Operators 
for Connection gauge fields }

Frame-like formalism  in higher spin field
theories, originally proposed by Vasiliev
and later developed in a number of works 
(e.g. see ~{\fvf, \fvf,
\vmaf, \vmas, \skvortsov, \skvortsovb, \vcubic, \boulskv,
\boulskvz})
is a powerful tool to describe gauge-invariant
interactions of higher spin fields in various
backgrounds including anti de Sitter (AdS) geometry.
Unlike the approach used by Fronsdal that considers
higher spin tensor fields as metric-type objects,
the frame-like formalism describes the higher spin dynamics
in terms of higher spin connection gauge fields
that generalize objects such as vielbeins and spin connections
in gravity (in standard  Cartan-Weyl formulation or
Mac Dowell-Mansoury-Stelle-West (MMSW) in case of nonzero
cosmological constant).
The higher spin connections  for a  given spin $s$ are 
described by collection
of two-row gauge fields
\eqn\grav{\eqalign{
\omega^{s-1|t}\equiv\omega_m^{a_1...a_{s-1}|b_1..b_t}(x)\cr
0\leq{t}\leq{s-1}\cr
1\leq{a,b,m}\leq{d}
}}
traceless in the fiber indices,
where $m$ is the curved $d$-dimensional space index while
$a,b$ label the tangent space with
 $\omega$ satisfying

\eqn\grav{\eqalign{\omega_m^{(a_1...a_{s-1}|b_1)..b_t}=0}}.

The gauge transformations for $\omega$
are given by

\eqn\grav{\eqalign{\omega_m^{a_1...a_{s-1}|b_1..b_t}
\rightarrow\omega_m^{a_1...a_{s-1}|b_1..b_t}+
D_m\rho^{a_1...a_{s-1}|b_1..b_t}}}
while the diffeomorphism symmetries
are
\eqn\grav{\eqalign{\omega_m^{a_1...a_{s-1}|b_1..b_t}(x)
\rightarrow\omega_m^{a_1...a_{s-1}|b_1..b_t}(x)+
\partial_m\epsilon^n(x)\omega_n^{a_1...a_{s-1}|b_1..b_t}(x)
\cr
+
\epsilon^n(x)\partial_n\omega_m^{a_1...a_{s-1}|b_1..b_t}(x)
}}
The $\omega^{s-1|t}$ gauge fields with $t\geq{0}$
are auxiliary fields related to the dynamical field
$\omega^{s-1|0}$ by generalized zero torsion constraints:
\eqn\lowen{\omega_m^{a_1...a_{s-1}|b_1...b_t}\sim
\partial^{b_1}...\partial^{b_t}{\omega_m^{a_1...a_{s-1}}}}
skipping pure gauge terms (for convenience of the notations,
we  set the cosmological constant to 1, anywhere the $AdS$ 
backgrounds are concerned)

It is also convenient to introduce the 
$d+1$-dimensional index
$A=(a,{\hat{d}})$ (where ${\hat{d}}$ labels the
extra dimension) and to 
combine $\omega^{s|t}$ into a single two-row field
$\omega^{A_1...A_{s-1}|B_1...B_{s-1}}(x)$
identifying
\eqn\grav{\eqalign{\omega^{s-1|t}=\omega^{a_1...a_{s-1}|b_1...b_t{\hat{d}}...
{\hat{d}}}
\cr
\omega^{A_1...A_{s-1}|B_1...B_{s-1}}V_{A_{t+1}}...V_{A_{s-1}}=
\omega^{A_1...A_{s-1}|B_1...B_{t}}
}}
where $V_A$ is the compensator field satisfying
$V_AV^A=1$.
The Fronsdal field $H^{a_1....a_s}$ is then obtained 
by symmetrizing
$\omega^{(a_1....a_s)}=e^{m(a_s}\omega_m^{a_1...a_{s-1})}$.
We now turn to the question of constructing vertex operators
for $\omega^{s-1|t}$.
The operators for spins greater than 2 constructed in 
our previous works
~{\spinself} were in fact limited to the Fronsdal-type objects
only.
In particular, in RNS superstring theory
the operators for $s=3$ are given by
\eqn\grav{\eqalign{V^{(-3)}=H_{abm}(p)
c{e^{-3\phi}}\partial{X^a}\partial{X^b}
\psi^me^{ipX}}}
at unintegrated minimal negative picture
and
\eqn\grav{\eqalign{V^{(+1)}=K\circ{H_{abm}(p)}\oint{dz}e^\phi
\partial{X^a}\partial{X^b}
\psi^me^{ipX}}}
at integrated minimal positive picture $+1$
where
$a,b,m=0,...{d-1}$ are Minkowski space-time indices,
$X^a(z)$ are space-time coordinates, $\psi^a$
are their worldsheet superpartners,
$b,c$ are reparametrizational fermionic ghosts
and $\beta,\gamma$ are bosonic superconformal ghosts.
 The homotopy transformation $K\circ{T}$
of an integrated operator
$T=\oint{dz}V(z)$  (with $V(z)$ being a primary field of dimension 1)
is defined according to
\eqn\grav{\eqalign{
K{\circ}T=T+{{(-1)^N}\over{N!}}
\oint{{dz}\over{2i\pi}}(z-w)^N:K\partial^N{W}:(z)
\cr
+{1\over{{N!}}}\oint{{dz}\over{2i\pi}}\partial_z^{N+1}{\lbrack}
(z-w)^N{K}(z)\rbrack{K}\lbrace{Q_{brst}},U\rbrace}}
where the BRST operator is
\eqn\lowen{Q=\oint{dz}\lbrace
cT-bc\partial{c}-{1\over2}\gamma\psi_m\partial
{X^m}-{1\over4}b\gamma^2\rbrace}
is the BRST operator,
$K=-4c{e^{2\chi-2\phi}}$ is homotopy operator satisfying
$\lbrace{Q,K}\rbrace=1$. $U$ and $W$ are the operators
appearing in the commutator
$\lbrack{Q,V(z)}\rbrack=\partial{U}(z)+W(z)$
and the ghost fields are bosonized as usual according to
\eqn\grav{\eqalign{
c=e^\sigma
\cr
b=e^{-\sigma}\cr
\gamma=e^{\phi-\chi}\equiv{e^\phi}\eta\cr
\beta=e^{\chi-\phi}\partial\chi\equiv{e^{-\phi}}\partial\xi
}}

The operators (7) and (8) are the elements of negative
and positive ghost cohomologies $H_{-3}$ and $H_1$
respectively (see ~{\spinself} for definitions and review).
They are related according to
$V^{(+1)}=:Z\Gamma^2{Z}\Gamma^2:V^{(-3)}$
by combination of BRST-invariant transformations
by picture-changing operators for $b-c$ and $\beta-\gamma$ systems:
$Z=:b\delta(T):$ and $\Gamma=:\delta(\beta)G:$
($T$ is the full stress tensor and $G$ is the supercurrent),
therefore the on-shell conditions and gauge transformations
for $H_{abm}$ at positive and negative pictures are identical.
The manifest expression  for $V^{(+1)}$ is given by
\eqn\lowen{
V_{s=3}(p;w)=\oint{{dz}}(z-w)^2U(z){\equiv}A_0+A_1+A_2+A_3+A_4+A_5+A_6+
A_7+A_8
}
where
\eqn\grav{\eqalign{
A_0(p;w)={1\over2}H_{abm}(p)\oint{dz}(z-w)^2
P^{(2)}_{2\phi-2\chi-\sigma}
e^{\phi}\partial{X^{a}}\partial{X^{b}}
\psi^{m}{e^{i{\vec{p}}{\vec{X}}}}(z)
}}
and
\eqn\grav{\eqalign{
A_8(w)=H_{abm}(p)\oint{dz}(z-w)^2
\partial{c}c\partial\xi\xi{e^{-\phi}}
\partial{X^{a}}\partial{X^{b}}
\psi^{m}\rbrace{e^{i{\vec{p}}{\vec{X}}}}(z)
}}
have ghost factors proportional to
$e^\phi$ and $\partial{c}c\partial\xi\xi{e^{-\phi}}$ respectively
and the rest of the terms carry ghost factor proportional to
$c\xi$:
\eqn\grav{\eqalign{
A_1(p;w)=-2H_{abm}(p)\oint{dz}(z-w)^2
c\xi({\vec{\psi}}\partial{\vec{X}})
\partial{X^{a}}\partial{X^{b}}
\psi^{m}
{e^{i{\vec{p}}{\vec{X}}}}(z)
\cr
A_2(p;w)=
-H_{abm}(p)\oint{dz}(z-w)^2c\xi
\partial{X^{a}}\partial{X^{b}}
\partial{X^{m}}P^{(1)}_{\phi-\chi}
{e^{i{\vec{p}}{\vec{X}}}}(z)
\cr
A_3(p;w)=
H_{abm}(p)\oint{dz}(z-w)^2c\xi
\partial{X^{a}}\partial{X^{b}}\partial^2{X^{m}}
{e^{i{\vec{p}}{\vec{X}}}}(z)
\cr
A_4(p;w)=
2H_{abm}(p)\oint{dz}(z-w)^2
c\xi\partial\psi^{a}P^{(1)}_{\phi-\chi}
\partial{X^b}\psi^{m}
{e^{i{\vec{p}}{\vec{X}}}}(z)
\cr
A_5(p;w)=
2H_{abm}(p)\oint{dz}(z-w)^2
c\xi\partial^2\psi^{a}
\partial{X^b}\psi^{m}
{e^{i{\vec{p}}{\vec{X}}}}(z)
\cr
A_5(p;w)=
-2H_{abm}(p)\oint{dz}(z-w)^2c\xi
\partial{X^{a}}\partial{X^{b}}(\partial^2{X^{m}}+
\partial{X^{a_3}}P^{(1)}_{\phi-\chi})
{e^{i{\vec{p}}{\vec{X}}}}(z)
\cr
A_6(p;w)=2iH_{abm}(p)\oint{dz}(z-w)^2
c\xi({\vec{p}}{\vec{\psi}})P^{(1)}_{\phi-\chi}
\partial{X^{a}}\partial{X^{b}}
\psi^{m}
{e^{i{\vec{p}}{\vec{X}}}}(z)
\cr
A_7(p;w)=
2iH_{abm}(p)\oint{dz}(z-w)^2
c\xi({\vec{p}}\partial{\vec{\psi}})
\partial{X^{a}}\partial{X^{b}}
\psi^{m}
{e^{i{\vec{p}}{\vec{X}}}}(z)
}}
Here $w$ is an arbitrary point in on the worldsheet
; since
all the $w$-derivatives of $s=3$ operators are BRST-exact in small
Hilbert space ~{\spinself}, all the correlation functions 
involving higher spin operators $V_{s=3}(p,w)$
are $w$-independent and the choice of $w$ is arbitrary.
Conformal dimension $n$ polynomials
$P^{(n)}_{A\phi+B\chi+C\sigma}$ (where
$A$, $B$, $C$ are some numbers) are defined according to
\eqn\lowen{e^{-A\phi(z)-B\chi(z)-C\sigma(z)}{{d^n}\over{dz^n}}
e^{A\phi(z)+B\chi(z)+C\sigma(z)}}
(where the product is understood
 in algebraic rather than
OPE sense).

As it is straightforward to check, the BRST-invariance constraints
on the operators (7) and (8)
lead to Pauli-Fierz type conditions
\eqn\grav{\eqalign{p^2H_{abm}=p^a{H_{abm}}
=\eta^{ab}H_{abm}=0}}
However, in general
\eqn\lowen{\eta^{am}H_{abm}\neq{0}}
as the 
 tracelessness in $a$ and $m$ or $b$ and $m$ indices
isn't required for $V^{(-3)}$ to be primary field.
In what follows below we shall interpret
$H_{abm}$ with the dynamical spin 3 connection form $\omega^{2|0}$,
identifying $m$ with the manifold index and $a,b$ with the fiber 
indices. So the tracelessness condition 
is generally imposed by BRST invariance
constraint on any pair  of fiber indices only
(but  not on a pair of manifold and fiber indices).
The same is actually true also for the
vertex operators for  frame-like gauge fields
  of spins higher than 3. Altogether, this corresponds
precisely to the double tracelessness constraints
for corresponding metric-like Fronsdal's fields for higher spins
(although the zero double trace condition does not
of course appear in the case of $s=3$)  
As  it is clear from the manifest expressions (7), (8)
the tensor $H_{abm}$ is by definition symmetric in indices
$a$ and $b$ and therefore can be represented as
a sum of two Young diagrams.
However, only the fully symmetric  diagram
is the physical state, since the second one
(with two rows) can be represented
as the BRST  commutator in the small Hilbert space:
\eqn\grav{\eqalign{
V^{(-3)}\sim\lbrace{Q,W}\rbrace\cr
W={H_{abm}}(p)c\partial\xi{e^{-4\phi+ipX}}
\partial{X^a}(\psi^{{\lbrack}m}\partial^2\psi^{b\rbrack}-
2\psi^{{\lbrack}m}\partial\psi^{b\rbrack}\partial\phi
\cr
+\psi^m\psi^b({5\over{13}}\partial^2\phi+{9\over{13}}(\partial\phi)^2))
+a\leftrightarrow{b}
}}
If $\Omega_{abm}$ is two-row,
the $V^{(-3)}$ operator is obtained
as the commutator of $W$ with the matter supercurrent
term of $Q$ given by $\sim\oint{\gamma}\psi_m\partial{X^m}$.
As $W$ commutes with $\oint
(-{1\over4}b\gamma^2-bc\partial{c})$ term in $Q$,
$V^{(-3)}$ is BRST-exact if and only if it commutes the stress
energy part of $Q$ given by $\oint{cT}$.
This is the case if the integrand of $W$ is a primary field.
It is, however, easy to check that the integrand 
is primary only when the last term
in its expression is present. Since this term is proportional to
$\sim\partial{\xi}e^{-4\phi+ipX}
\partial{X^a}\psi^m\psi^b
({5\over{13}}\partial^2\phi+{9\over{13}}(\partial\phi)^2)$
it is automatically antisymmetric in $m$ and $b$.
and is absent when multiplied by fully symmetric $H_{abm}$.
In the latter case this term is not a primary since its OPE
with $T$ contains cubic singularities
and therefore the commutator of $Q$ with $W$ does not give
$V^{(3)}$.
Similarly, shifting $H_{abm}$ by symmetrized derivative
$H_{abm}\rightarrow{H_{abm}}+p_{(m}\Lambda_{ab)}$
is equivalent to shifting the vertex operator (7) by
BRST exact terms given by
\eqn\grav{\eqalign{V^{(-3)}\rightarrow{V^{(-3)}+\lbrace{Q,U}\rbrace}
\cr
U=\Lambda_{ab}
c\partial\xi{e^{-4\phi+ipX}}\lbrace
\partial{X^a}((p\psi)\partial^2\psi^{b}-
2(p\psi)\partial\psi^{b}\partial\phi
\cr
+(p\psi)\psi^b({5\over{13}}\partial^2\phi
+{9\over{13}}(\partial\phi)^2))
\cr
+\partial{X^a}\partial{X^b}((p\partial^2{X})-
\partial\phi(p\partial{X}))\rbrace
}}
Of course everything described above also applies
to the vertex operator (8) at positive picture, with
appropriate $Z,\Gamma$ transformations.
This altogether already sends a strong hint to relate
(7), (8)  to vertex operators
for the dynamical frame-like field $\omega^{2|0}$
describing spin 3.
However, to make the relation  between string theory
and frame-like
formalism, we still need the vertex operators for
the remaining extra fields $\omega^{2|1}$ and $\omega^{2|2}$.
The expressions that we propose are given by
\eqn\grav{\eqalign{
V^{2|1}(p)=2\omega_m^{ab|c}(p)ce^{-4\phi}(
-2\partial\psi^m\psi_c\partial{X_{(a}}\partial^2{X_{b)}}
\cr
-2\partial\psi^m\partial\psi_c\partial{X_a}\partial{X_b}
+\psi^{m}\partial^2\psi_{c}\partial{X_a}\partial{X_b})e^{ipX}
}}
for $\omega^{2|1}$ and
\eqn\grav{\eqalign{
V^{2|2}(p)=-3\omega_m^{ab|cd}(p)
ce^{-5\phi}(\psi^{m}\partial^2\psi_c\partial
^3\psi_{d}\partial{X^a}\partial{X_b}
-2
\psi^{m}\partial\psi_c\partial
^3\psi_{d}\partial{X_{a}}\partial^2{X_{b}}
\cr
+{5\over8}\psi^{m}\partial\psi_c\partial
^2\psi_{d}\partial{X_{a}}\partial^3{X_{b}}
+{{57}\over{16}}
\psi^{m}\partial\psi_c\partial
^2\psi_{d}\partial^2{X_{a}}\partial^2{X_{b}})e^{ipX}
}}
for $\omega^{2|2}$.
We start with analyzing the operator for $\omega^{2|1}$
Straightforward application of $\Gamma$ to this
operator gives 
\eqn\grav{\eqalign{:
:\Gamma{V^{2|1}}:(p)=V^{(-3)}(p)\cr
H_{m}^{ab}(p)=ip_c\omega_m^{ab|c}(p)}}
i.e. the picture-changing of $V^{2|1}$ gives
the vertex operator for $\omega^{2|0}$ with the
3-tensor given by the divergence of $\omega^{2|1}$,
i.e. for $p_c\omega_m^{ab|c}(p)\neq{0}$ $V^{2|1}$
is the element of $H_{-3}$. If,however, the divergence 
vanishes, the cohomology rank changes and 
$V^{2|1}$ shifts to $H_{-4}$. This is precisely the
case we are interested in.
Namely, consider the $H_{-4}$ cohomology condition
\eqn\lowen{p_c\omega_m^{ab|c}(p)={0}}
The general solution of this constraint is
\eqn\grav{\eqalign{\omega_m^{ab|c}
=2p^c\omega_m^{ab}-p^a\omega_m^{bc}
-p^b\omega_m^{ac}+p_d\omega_m^{acd;b}}}
where $\omega_m^{ab}$ is traceless and divergence free
in $a$ and $b$ and satisfies the same on-shell constraints
as $H_m^{ab}$, while
$\omega_m^{acd;b}$ is some three-row field,
antisymmetric in $a,c,d$ and symmetric in $a$ and $b$.
It is, however, straightforward to check that
the operator $V^{2|1}$ with the polarization given by
$\omega^{ab|c}=p_d\omega_m^{acd;b}$ can be cast as
the BRST commutator:
\eqn\grav{\eqalign{
p_d\omega_m^{acd;b}(p)V^m_{ac|b}(p)
=\lbrace{Q},
\omega_m^{acd;b}(p)\oint{dz}e^{\chi-5\phi+ipX}\partial\chi
(
-2\partial\psi^m\psi_c\partial{X_{a}}\partial^2{X_{b}}
\cr
-2\partial\psi^m\partial\psi_c\partial{X_a}\partial{X_b}
+\psi^{m}\partial^2\psi_{c}\partial{X_a}\partial{X_b})
\cr\times
(\partial^2\psi_d-{4\over3}\partial\psi_d\partial\phi
+{1\over{141}}\psi_d(41(\partial\phi)^2-29\partial^2\phi))
\rbrace
}}

therefore, modulo pure gauge terms
the cohomology condition (24) is 
the zero torsion condition relating the extra field
$\omega^{2|1}$ to the dynamical $\omega^{2|0}$ connection.
Similarly, constraining $V^{2|2}$ to be the element
of $H_{-5}$ cohomology results in
the second generalized zero torsion condition
\eqn\grav{\eqalign{
\omega_m^{ab|cd}=2p^d\omega^{ab|c}-p^a\omega^{bd|c}
-p^b\omega^{ad|c}+2p^c\omega^{ab|d}-p^a\omega^{bc|d}
-p^b\omega^{ac|d}}}
relating $\omega^{2|2}$ to $\omega^{2|1}$ modulo
BRST-exact terms $\sim\lbrace{Q,W^{2|2}(p)}\rbrace$ where
\eqn\grav{\eqalign{W^{2|2}(p)
=\omega^{ab;cdf}(p)\oint{dz}e^{ipX}
{\lbrack}(\psi^{m}\partial^2\psi_c\partial
^3\psi_{d}\partial{X^a}\partial{X_b}
-2
\psi^{m}\partial\psi_c\partial
^3\psi_{d}\partial{X_{a}}\partial^2{X_{b}}
\cr
+{5\over8}\psi^{(m}\partial\psi_c\partial
^2\psi_{d)}\partial{X_{a}}\partial^3{X_{b}}
+{{57}\over{16}}
\psi^{m}\partial\psi_c\partial
^2\psi_{d}\partial^2{X_{a}}\partial^2{X_{b}})
\rbrack\cr
\times
(-{5\over2}L_f\partial^2\xi+\partial{L_f}\partial\xi)}}
where, as previously, $\xi=e^{\chi}$ and
\eqn\grav{\eqalign{
L_f=e^{-6\phi}(\partial^2\psi_f-\partial\psi_f\partial\phi
+{3\over{25}}\psi_f((\partial\phi)^2-4\partial^2\phi))}}.
The gauge transformations for $\omega^{2|1}$ and $\omega^{2|2}$:
$\delta\omega_m^{ab|c}(p)=p_m\Lambda^{ab|c}$
and 
$\delta\omega_m^{ab|cd}(p)=p_m\Lambda^{ab|cd}$
with $\Lambda$'s having having the same symmetries
in the fiber indices as $\omega$'s shift
the operators (21), (22) by terms that
are BRST-exact in the small Hilbert space;
the explicit expressions for the appropriate
BRST commutators are given in the Appendix B.
Similarly to the $\omega^{2|0}$ case,
for $\omega^{2|1}$ and $\omega^{2|2}$
with the manifold $m$ index antisymmetric with any of
the fiber indices $a$ or $b$
the operators (21), (22) become BRST-exact in the small
Hilbert space. Given the cohomology
(``zero torsion'') conditions (25), (27) 
 ensures that the fully symmetric
symmetric $s=3$ Fronsdal field is related to
dynamical field $\omega^{2|0}$
 by the gauge transformation removing  the 
two-row diagram. The expressions for the appropriate
BRST commutators are given in the Appendix B.

This concludes the construction of the vertex operators 
for frame-like gauge fields for spin 3.
In the next section we shall use this construction
to analyze the 3-point open string amplitude for spin 3.

\centerline{\bf 3. 3-Point Amplitude and 3-Derivative Vertex}
In this section we use the vertex operator formalism, developed
in the previous section, to compute the cubic coupling
of massless spin 3 fields. In this paper we limit ourselves
to the 3-derivative contributions corresponding to the Berends,
Burgers and Van Dam (BBD)
~{\bbd, \boulanger} type
 vertex in the field theory limit.
The first  step is to choose the ghost pictures of the operators to
 ensure the correct ghost number balance, i.e.
so that the correlator has total $\phi$-ghost number $-2$,
$b-c$ ghost number $+3$ and $\chi$-ghost number $+1$. 
This requires
two out of three operators to be taken unintegrated at negative
pictures and the third one at positive picture (note that higher
spin operators at positive pictures are always integrated).
It is convenient to take unintegrated operators at the
minimal ghost picture $-3$, i.e. we shall use the 
$V^{(-3)}$ operator for $\omega^{2|0}$. Then the remaining integrated
operator must be taken at picture $+5$, and
only the terms proportional to the ghost factor ${\sim}c{e^{\chi+4\phi}}$
will contribute, while the terms
proportional to $\sim{\partial{c}}c{e^{2\chi+3\phi}}$ and
to ${\sim}e^{5\phi}$ will drop out as they don't satisfy the balance
of ghosts. It is therefore appropriate to choose the operator
for $\omega^{2|2}$ for the third operator (which minimal positive picture
is $+3$) and to apply the picture changing transformation
twice to bring it to the picture $+5$.
The result is given by
\eqn\grav{\eqalign{
\lbrace{Q,\xi\lbrace{Q},\xi{K}\circ\omega_m^{ab|cd}(p)\oint{d}e^{3\phi}
F^{m({{17}\over2})}_{abcd}}\rbrace\rbrace=V_1(p)+V_2(p)
\cr
\equiv\omega_m^{ab|cd}(p)\oint{du}(u_0-u)^8ce^{\chi+4\phi+ipX}R^m_{ab|cd}(u)
}}
where
\eqn\grav{\eqalign{V_1(p)
={3\over{64}}\omega_m^{ab|cd}(p)\oint{du}(u_0-u)^8ce^{\chi+4\phi+ipX}
L^{m({{9}})}_{abcd}
\cr\times\lbrace
{{24}\over{11!}}P^{(11)}_{2\phi-2\chi-\sigma}({1\over{8}}P_{\chi}^{(2)}
+{1\over{8}}P^{(2)}_{2\phi-2\chi-\sigma}
-{1\over4}P^{(1)}_{\chi}P^{(1)}_{2\phi-2\chi-\sigma}
\cr
-12P^{(11)}_{\phi-\chi}P^{(1)}_{2\phi-2\chi-\sigma}
-12(P^{(1)}_{\phi-\chi})^2)
\cr
+{1\over{11!}}P^{(12)}_{2\phi-2\chi-\sigma}
P^{(1)}_{-{3\over{2}}\phi+{{55}\over{4}}\chi+{{11}\over{4}}\sigma}
-{{102}\over{13!}}P^{(13)}_{2\phi-2\chi-\sigma}\rbrace}}
and
\eqn\grav{\eqalign{V_2(p)=
{1\over{9!-8!}}\sum_{n=0}^7{2^{n-7}}
\sum_{{\lbrace}l,m,p,q\geq{0};l+m+p+q=8-n\rbrace}
\sum_{r=0}^p\sum_{a=0}^l\sum_{b=0}^q\sum_{N=0}^{a+b+r+5}
\cr{\lbrace}
{{(-1)^{a+b+p+q}N!}\over{m!l!(p-r)!r!(N-r)!(5+a+b+r-N)!}}
\cr\times
\omega_m^{ab|cd}(p)\oint{du}(u_0-u)^8ce^{\chi+4\phi}
\partial^{(p-r)}L^{m({{N+9}})}_{abcd}
\cr\times
\partial^{(m)}P^{n|8}_{2\phi-2\chi-\sigma|\chi}
P^{l-a|l}_{\chi|\phi-\chi}P^{q-b|q}_{3\phi+\chi|\phi-\chi}
P^{(5+a+b+r-N)}_{\phi-\chi}(u)\rbrace}}

Here $u_0$ is an arbitrary point on the boundary (will be fixed later)
and

\eqn\grav{\eqalign{
F^{m({{17}\over2})}_{abcd}=
(\psi^{m}\partial^2\psi_c\partial
^3\psi_{d}\partial{X^a}\partial{X_b}
-2
\psi^{m}\partial\psi_c\partial
^3\psi_{d}\partial{X_{a}}\partial^2{X_{b}}
\cr
+{5\over8}\psi^{m}\partial\psi_c\partial
^2\psi_{d}\partial{X_{a}}\partial^3{X_{b}}
+{{57}\over{16}}
\psi^{m}\partial\psi_c\partial
^2\psi_{d}\partial^2{X_{a}}\partial^2{X_{b}})
e^{ipX}}}
is dimension ${{17}\over2}$ primary field
(given the on-shell conditions on $\omega$);
conformal dimension $N+9$
fields  are defined as the OPE terms in the product of
$F^{m({{17}\over2})}_{abcd}$ with the matter supercurrent
$G=-{1\over2}\psi_n\partial{X^n}$ on the worldsheet:
\eqn\grav{\eqalign{G_m(z)
F^{m({{17}\over2})}_{abcd}(w)=\sum_{N=0}^\infty(z-w)^{N-1}
L^{m({{N+9}})}_{abcd}(w)}}
or manifestly
\eqn\grav{\eqalign{
L^{m({{N+9}})}_{abcd}
={{e^{ipX}}\over{N!}}\lbrace{{15}\over8}
(\partial{X_a}\partial^3{X_b}+{{171}\over{16}}
\partial^2{X_a}\partial^2{X_b})(
\partial^{(N+1)}X^m\partial\psi_c\partial^2\psi_d
\cr
-{1\over{{N+1}}}\partial^{(N+2)}X_c\psi^m\partial^2\psi_d
+{{2}\over{(N+1)(N+2)}}\partial^{(N+3)}X_d\psi^m\partial\psi_c
)
\cr
+
\psi^m\partial\psi_c\partial^2\psi_d(-{{15}\over{{8(N+1)}}}
\partial^{N+1}\psi_a\partial^3{X_b}
-{{45}\over{{4(N+1)(N+2)(N+3)}}}
\partial^{N+3}\psi_b\partial{X_a}
\cr
-{{171}\over{{8(N+1)(N+2)}}}
\partial^{N+2}\psi_a\partial^2{X_b})
\cr
+3\partial{X_a}\partial{X_b}(\partial^{(N+1)}X^m\partial^2\psi_c
\partial^3\psi_d
\cr
-{{2}\over{{(N+1)(N+2)}}}\partial^{(N+3)}X_c
\psi^m
\partial^3\psi_d
-
{{6}\over{{(N+1)(N+2)(N+3)}}}
\partial^{(N+4)}X_d\partial^2\psi_c
\psi^m
\cr
-{3\over{N+1}}\psi^m\partial^2\psi_c\partial^3\psi_d\partial^{(N+1)}
\psi_{(a}\partial{X_b)})
\cr
-6\partial{X_a}\partial^2{X_b}
(\partial^{(N+1)}X^m\partial\psi_c
\partial^3\psi_d
\cr
-{{1}\over{{N+1}}}\partial^{(N+2)}X_c
\psi^m
\partial^3\psi_d
-
{{6}\over{{(N+1)(N+2)(N+3)}}}
\partial^{(N+4)}X_d\partial\psi_c
\psi^m
)
\cr
+6\psi^m\partial\psi_c\partial^3{X_d}
({1\over{N+1}}\partial^{(N+1)}\psi_a\partial^2{X_b}
+
{{2}\over{{(N+1)(N+2)}}}
\partial^{(N+2)}\psi_b\partial{X_a})
\cr+N:\partial^{N-1}G_mF^{m({{17}\over2})}_{abcd}:(1-\delta_{0;N})
-i:(p^n\partial^N\psi_n)F^{m({{17}\over2})}_{abcd}:
\rbrace
}}
The associate ghost polynomials
$P^{n|N}_{A_1\phi+B_1\chi+C_1\sigma|A_2\phi+B_2\chi+C_2\sigma}$
are conformal dimension $n$ polynomials
in derivatives of $\phi$,$\chi$ and $\sigma$
defined as the terms in the operator product
\eqn\grav{\eqalign{
P^{(N)}_{A_1\phi+B_1\chi+C_1\sigma}(z)e^{A_2\phi+B_2\chi+C_2\sigma}(w)
\cr
=\sum_{n=0}^N(z-w)^{n-N}
P^{n|N}_{A_1\phi+B_1\chi+C_1\sigma|A_2\phi+B_2\chi+C_2\sigma}(z)
e^{A_2\phi+B_2\chi+C_2\sigma}(w)}}
(see the Appendix A for some of the 
techniques related to these polynomials).
Note that, for example,
\eqn\grav{\eqalign{P^{N|N}_{A_1\phi+B_1\chi+C_1\sigma|A_2\phi+B_2\chi+C_2\sigma}
\equiv{P^{(N)}_{A_1\phi+B_1\chi+C_1\sigma}}}}
while
\eqn\grav{\eqalign{P^{0|N}_{A_1\phi+B_1\chi+C_1\sigma|A_2\phi+B_2\chi+C_2\sigma}
=\prod_{k=0}^{n-1}(C_1C_2+B_1B_2-A_1A_2-k)}}

We are now prepared  to analyze the 3-point function given by
\eqn\grav{\eqalign{A(p,k,q)=
\omega_n^{s_1s_2}(p)\omega_p^{t_1t_2}(k)\omega_m^{ab|cd}(q)
\oint{du}(u-u_0)^8
\cr\times
<c\partial{X_{s_1}}\partial{X_{s_2}}\psi^n{e^{ipX}}(z)
c\partial{X_{t_1}}\partial{X_{t_2}}\psi^n{e^{ikX}}(w)
ce^{\chi+4\phi+iqX}R^m_{ab|cd}(u)>}}
Using the $SL(2,R)$ symmetry, it is convenient to
set $z\rightarrow\infty,u_0=w=0$ (see ~{\spinselff} where the
details related to this choice were discussed). 
For the notation purposes, however, it is convenient
to retain $z$ and $w$ in our notations for a time being.
We start with computing the ``static''  exponential ghost
part of the correlator.
Simple calculation gives
\eqn\grav{\eqalign{<ce^{-3\phi}(z)ce^{-3\phi}(w)ce^{4\phi+\chi(u)}>
\cr
=(z-w)^{-8}(z-u)^{13}(w-u)^{13}{\rightarrow}z^5(w-u)^{13}}}.
where we substituted the $z\rightarrow\infty$ limit.
Next, consider the $\psi$-part of the correlator.
The expression for $R^m_{ab|cd}$ contains two types of terms:
those that are quadratic in $\psi$
and those that are quartic $\psi$ .
Since the remaining two spin 3 operators
are linear in $\psi$, only the quadratic terms contribute to the 
correlator.
Note that all the terms quadratic in $\psi$ are also cubic in 
$\partial{X}$. So the pattern for the $\psi$-correlators
is
\eqn\grav{\eqalign{
<\psi^n(z)\psi^p(w):\partial^{(P_1)}\psi_c\partial^{(P_2)}\psi_d:(u)>
\cr
=P_1!P_2!({{\eta_d^n\eta_c^p}\over{z^{P_2+1}(w-u)^{P_1+1}}}
-{{\eta_c^n\eta_d^p}\over{z^{P_1+1}(w-u)^{P_2+1}}})}}
where, according to the manifest expression (35) for
$R^m_{ab|cd}$ the numbers $P_1$ and $P_2$ can vary from $0$
to $N+3$ (and $N_{max}=8$).
Next, consider the $X$-part. As in this paper we limit ourselves to
just three-derivative terms, it is sufficient to compute the terms linear
in momentum (since the $\omega^{2|2}$ field already contains
2 derivatives out of 3).
According  to (31) , (32) and (35)
 the $X$-factor is a combination of the 3-point
correlators of the type 
${\sim}(\omega^{2|0})^2\omega^{2|2}
<(\partial{X})^2{e^{ipX}}(z)(\partial{X})^2{e^{ikX}}(w)
\partial^{(M_1)}X\partial^{(M_2)}X\partial^{(M_3)}X{e^{iqX}}>$
with different values of $M_1$, $M_2$ and $M_3$.
Straightforward computation gives
\eqn\grav{\eqalign{
lim_{z\rightarrow\infty}
\omega_n^{s_1s_2}(p)\omega_p^{t_1t_2}(k)
\omega_m^{ab|cd}(q)
\cr\times
<\partial{X_{s_1}}\partial{X_{s_2}}e^{ipX}(z)
\partial{X_{t_1}}\partial{X_{t_2}}e^{ikX}(w)
\partial^{(M_1)}{X_{a}}\partial^{(M_2)}{X_{b}}
\partial^{(M_1)}{X^{m}}e^{iqX}(u)>
\cr
=M_1!M_2!M_3!\omega_n^{s_1s_2}(p)\omega_p^{t_1t_2}(k)
\omega_m^{ab|cd}(q)
\lbrace
{{2iq_{t_2}\eta_{s_1a}\eta_{s_2b}
\eta_{t_1}^m}\over{z^{2+M_1+M_2}(w-u)^{2+M_3}}}
\cr
+
iq_{t_2}\eta_{s_1a}\eta_{s_2}^m\eta_{t_1b}
({1\over{z^{2+M_1+M_3}(w-u)^{2+M_2}}}+
{1\over{z^{2+M_2+M_3}(w-u)^{2+M_1}}})
\cr
-
{{2ik_{s_2}\eta_{t_1a}\eta_{t_2b}
\eta_{s_1}^m}\over{z^{3+M_3}(w-u)^{1+M_1+M_2}}}
\cr
-
ik_{s_2}\eta_{t_1a}\eta_{t_2}^m\eta_{s_1b}
({1\over{z^{3+M_1}(w-u)^{1+M_2+M_3}}}+
{1\over{z^{3+M_2}(w-u)^{1+M_1+M_3}}})
\cr
-{1\over{M_3}}\eta_{s_1t_1}
\eta_{s_2a}\eta_{t_2b}
(ik^m({1\over{z^{3+M_1}(w-u)^{1+M_2+M_3}}}+
{1\over{z^{3+M_2}(w-u)^{1+M_1+M_3}}})
\cr
+ip^m(
{1\over{z^{3+M_1+M_3}(w-u)^{1+M_2}}}+
{1\over{z^{3+M_2+M_3}(w-u)^{1+M_1}}}
))
\cr
+ip_b\eta_{s_1t_1}\eta_{s_2a}\eta_{t_2}^m
({1\over{M_2}}{1\over{z^{3+M_1}(w-u)^{1+M_2+M_3}}}+
{1\over{M_1}}{1\over{z^{3+M_2}(w-u)^{1+M_1+M_3}}})
\cr
+{{ip_b\eta_{s_1t_1}\eta_{s_2}^m\eta_{t_2a}}\over
{z^{3+M_3}(w-u)^{1+M_1+M_2}}}({1\over{M_1}}+{1\over{M_2}})
}}
Comparing this with the explicit
expression (30)-(35) for the $\omega^{2|2}$ vertex operator, 
it is easy to notice that, while the static
ghost factor (40) is proportional to $\sim{z^5}$,
the $z$-asymptotics of the $\psi$-correlator (41)
 is $\sim{1\over{z}}+ O({1\over{z^2}})$ and
the asymptotics for the $X$-correlator
is $\sim{1\over{z^4}}+O({1\over{z^5}})$.
This means that only the terms
proportional to $\sim{z^0}$ contribute to the interaction
vertex. Terms proportional to negative powers of $z$ disappear
in the limit $z\rightarrow\infty$  and correspond to
pure gauge contributions.
There are no terms proportional to positive powers
of $z$ (their presence would be a signal of problems
with the gauge invariance).
Moreover the $z$-asymptotics further simplifies the analysis
of the ghost polynomials in the expressions (30)-(32) for
the $\omega^{2|2}$ operator; namely, all the polynomials
have to couple to the $c{e^{-3\phi}}$ ghost exponent
of the $\omega^{2|0}$ operator  sitting at $w$ as any couplings
of these polynomials with the operator sitting at $z$ produce
contributions vanishing in the $z\rightarrow\infty$ limit.

Combining (30), (32), (41 and (42) 
we arrive to the following expression
for the main matter building block for the matter
part of the correlator:
\eqn\grav{\eqalign{
lim_{z\rightarrow\infty}
\omega_n^{s_1s_2}(p)\omega_p^{t_1t_2}(k)
\omega_m^{ab|cd}(q)
\cr\times
<\psi^n\partial{X_{s_1}}\partial{X_{s_2}}e^{ipX}(z)
\psi^p\partial{X_{t_1}}\partial{X_{t_2}}e^{ikX}(w)
L^{m({{N+9}})}_{abcd}(u;q)>
\cr
=
z^{-5}(w-u)^{-9-N}A_N(p,k,q)\cr
A_N(k,p,q)=
\omega_n^{s_1s_2}(p)\omega_p^{t_1t_2}(k)
\omega_m^{ab|cd}(q)\times\lbrace
\eta^{nm}\eta_{pd}(-72(N+5)\eta^{s_1a}\eta^{s_2b}\eta^{t_1c}q^{t_2}
\cr
+(72(N+5)-{{45}\over{4}}(N+1)^2(N+2)-144)
\eta^{t_1a}\eta^{s_1b}\eta^{t_2c}k^{s_2}
\cr
+(144-{{45}\over{4}}N(N+1))
\eta^{s_1t_1}\eta^{s_2a}\eta^{t_2b}k^{c}
\cr
+
({{45}\over{4}}N(N+1)^2-72(N+4))
\eta^{s_1t_1}\eta^{s_2a}\eta^{t_2c}p^{b}
)+Symm(m,a,b)\rbrace
}}
Using the manifest expression
(35) for the $V_{2|2}$ vertex operator in terms of
$L_{9+N}$ and their derivatives, it is now straightforward
to calculate the cubic coupling.
First of all, it is immediately clear that only
the $V_2$-part of $V_{2|2}$ contributes to the
overall correlator.
No terms from $V_1$ contribute since, as it was pointed
out above, all the ghost
polynomials entering $V_{2|2}$ must be completely absorbed
by the ghost exponent $\sim{ce^{-3\phi}}$ located at $w$
(no couplings to the exponent at $z$ are allowed as 
they would result in contributions vanishing at $z\rightarrow\infty$).
At the same time all the terms in $V_1$  carry the factors
of $P^{(n)}_{2\phi-2\chi-\sigma}$ ($n=11,12,13$) which cannot
be absorbed by $ce^{-3\phi}$ (i.e. their OPE's with $ce^{-3\phi}$)
are less singular than $(z-w)^{-n}$)
Indeed, since
$e^{2\phi-2\chi}b(z)ce^{-3\phi}(w)\sim{(z-w)^5{e^{-\phi-2\chi}}(w)+
O(z-w)^6}$, clearly for $n\geq{5}$
\eqn\grav{\eqalign{
\partial^{(n)}(e^{2\phi-2\chi}b)(z)\equiv{c}e^{-3\phi}(w)
\equiv:P^{(n)}_{2\phi-2\chi-\sigma}e^{2\phi-2\chi}b:{c}e^{-3\phi}(w)
\cr
\sim{{n!}\over{(n-5)!}}:P^{(n-5)}_{2\phi-2\chi-\sigma}{e^{-\phi-2\chi}}:(w)
+O(z-w)}}
implying that
\eqn\grav{\eqalign{
P^{(n)}_{2\phi-2\chi-\sigma}(z){c}e^{-3\phi}(w)\sim{O}({1\over{z^5}})}}
i.e. no complete contractions for $n\geq{6}$.
Next, combining (42) with the expression (35) for $V_2(q)$ we obtain
the following result 
for the overall correlator:
\eqn\grav{\eqalign{
{6\over{9!-8!}}\sum_{n=0}^7{2^{n-7}}
\sum_{{\lbrace}l,m,p,q\geq{0};l+m+p+q=8-n\rbrace}
\sum_{r=0}^p\sum_{b=0}^q\sum_{N=l+b+r+2}^{l+b+r+5}
\cr
\lbrace
{{(-1)^{1+b+m+q+r+N}N!(N+8+p-r)!\prod_{j=0}^{m-1}(n+j)!}
\over{m!l!(p-r)!r!(N-r)!(5+l+b+r-N)!(N-l-b-r-2)!(N+8)!}}
\cr\times
\alpha_{3;1;0|1;-1;0}^{-3;0;1}(q-b|q)
\alpha_{2;-2;-1|0;1;0}^{-3;0;1}(n|8)A_{N}(p,k,q)\rbrace
}}

where we used the fact that the only
nonzero contributions from the summation 
over $a$ are the terms with $a=l$ 
for which $P^{l-a|l}_{\chi|\phi-\chi}=P^{0|l}_{\chi|\phi-\chi}=
(-1)^l{l!}$; while for $a\neq{l}$  $P^{l-a|l}_{\chi|\phi-\chi}$
are the polynomials in $\chi$ of dimension $l-a$  which
aren't contractible with $ce^{-3\phi}$.
The numbers $\alpha_{A_1;B_1;C_1|A_2;B_2;C_2}^{A_3;B_3;C_3}(n|N)$
appearing in (46) are the coefficients in front of the leading
order
 terms in the operator products 
\eqn\grav{\eqalign{
P^{n|N}_{A_1\phi+B_1\chi+C_1\sigma|A_2\phi+B_2\chi+C_2\sigma}(z)
e^{A_3\phi+B_3\chi+C_3\sigma}(w)
\cr
\sim
{{\alpha_{A_1;B_1;C_1|A_2;B_2;C_2}^{A_3;B_3;C_3}(n|N)}\over{(z-w)^n}}
e^{A_3\phi+B_3\chi+C_3\sigma}(w)}}
The calculation of these coefficients is explained
and the values are given in the Appendix A (see (58), (60)).
Finally, substituting for $A_N(p,k,q)$ and
evaluating the series (46) we obtain the following answer
for the cubic coupling:
\eqn\grav{\eqalign{ A(p,k,q)=
{{691072283467i}\over{720}}
\omega_n^{s_1s_2}(p)\omega_p^{t_1t_2}(k)
\omega_m^{ab|cd}(q)\times\lbrace
\eta^{nm}\eta_{pd}({1\over{36}}\eta^{s_1a}\eta^{s_2b}\eta^{t_1c}q^{t_2}
\cr
+{4\over3}
\eta^{t_1a}\eta^{s_1b}\eta^{t_2c}k^{s_2}
\cr
+{1\over{12}}
\eta^{s_1t_1}\eta^{s_2a}\eta^{t_2b}k^{c}
\cr
-
\eta^{s_1t_1}\eta^{s_2a}\eta^{t_2c}p^{b}
)+Symm(m,a,b)\rbrace
}}

This concludes the calculation of the 3-derivative part
of the cubic vertex.
Inclusion of the appropriate Chan-Paton's indices is straightforward
and leads to vertices of the type considered in ~{\boulanger, \bbd}.

\centerline{\bf 4. Conclusion and Discussion}

In this paper we performed  the analysis of open string
vertex operators describing generalized connection gauge
fields in Vasiliev's frame-like formalism for higher spin fields.
We have shown that generalized zero curvature conditions
relating
auxiliary connections $\omega^{s-1|t}$  to
the dynamical $\omega^{s|0}$ fields 
are realized (up to BRST-exact terms) through ghost cohomology
conditions on vertex operators that ensure that
the fields with higher values of $t$ belong to cohomologies
of higher orders.
We have also given precise BRST arguments 
relating $\omega^{2|0}$ to symmetric Fronsdal 
field for spin 3, presenting BRST commutator 
for non-symmetric
spin 3 diagram
 (an important point which has been
somewhat obscure before). 
We also demonstrated
how the 3-derivative cubic vertex of spin 3 fields
appears from string-theoretic 3-point amplitude.
computed in this work.
Obvious directions for future research include
the computation of the 5-derivative  vertex in the flat space
(which technically appears to be significantly more tedious
than the 3-derivative one) and to generalize the construction
proposed in this work to frame-like gauge fields with
spins greater than 3. We hope to present these results
soon in our future papers.
The cubic vertex  computed in this work is the one for the flat space
and an important next step would be to generalize it to AdS.
For that, one has to generalize the computation, analyzing 
of the 3-point function of operators for frame-like spin 3 fields 
in the sigma-model background studied in ~{\selfsigma}.
That is, one has to perturb the flat background
with the vertex operators for spin 2 vielbeins and connections
in AdS space constructed in ~{\selfsigma, \sjprogress}. 
These operators carry negative
cosmological constant and the vacuum solution of the low-energy
equations of motion is described by AdS geometry.
To calculate the cubic coupling of spin 3 frame-like fields in the 
AdS space one has to consider their disc amplitude
with insertions of closed string operators for spin 2 connections.
As the insertions carry the dependence on the cosmological
constant parameter,
 important question to explore is
the relation of this amplitude with to AdS deformations
of flat vertices considered by Vasiliev by methods
of vertex complex analysis ~{\vcubic}. It is particularly
interesting to clarify how the insertions of the closed string
operators give rise to terms
with lower the number of derivatives, as it was observed
 in ~{\vcubic} for the AdS deformations of vertices in Minkowski space.
Another issue is to explore the relevance of the zero
momentum parts of the vertex operators for frame-like fields
to space-time symmetry generators and higher spin algebra
in AdS space. Typically, physical vertex operators
in string theory are related to generators of global space-time
symmetries in the zero momentum limit. For example,
a photon operator at $p=0$ is the generator of translations.
The similar question can be asked about the vertex operators
of frame-like gauge fields for higher spins at zero momentum.
While in general these operators at $p=0$ do not generate
global symmetries for RNS string theory in flat space,
it is possible that they realize the symmetries
of the sigma model perturbed by operators
for spin 2  connections and vielbeins, provided that
AdS vacuum constraints are imposed on spin 2.
So far we have been able to show this for spin 3 only 
~{\meinprogress} and
this conjecture needs to be generalized for 
higher spins. If the vertex operators for the frame-like
gauge fields are  indeed related to the symmetries of the sigma-model,
their operator algebras may provide  nice realizations of
higher  spin algebras in various AdS backgrounds.

\centerline{\bf Acknowledgements}

One of the authors (D.P.) acknowledges useful discussions
with R. Metsaev, S.-J. Rey, E. Skvortsov and M. Vasiliev.

\vfill\eject

\centerline{\bf Appendix A. Associate Ghost Polynomials and 
$\alpha_{A_1;B_1;C_1|A_2;B_2;C_2}^{A_3;B_3;C_3}(n|N)$-coefficients}

In this Appendix section we explain some of the techniques
to calculate the $\alpha$-coefficients that appear in the 
series (46) for the spin 3 cubic coupling.
As was explained above, the  associate ghost polynomials
 $P^{n|N}_{A_1\phi+B_1\chi+C_1\sigma|A_2\phi+B_2\chi+C_2\sigma}$
$(0\leq{n}\leq{N})$
are the conformal dimension $n$ polynomials in bosonized
ghost fields $\phi,\chi$ and $\sigma$, defined as the OPE terms
in the product

\eqn\grav{\eqalign{
P^{(N)}_{A_1\phi+B_1\chi+C_1\sigma}(z)e^{A_2\phi+B_2\chi+C_2\sigma}(w)
\cr
=\sum_{n=0}^N(z-w)^{n-N}
P^{n|N}_{A_1\phi+B_1\chi+C_1\sigma|A_2\phi+B_2\chi+C_2\sigma}(z)
e^{A_2\phi+B_2\chi+C_2\sigma}(w)}}

where the conformal dimension $N$ polynomials 
$P^{(N)}_{A_1\phi+B_1\chi+C_1\sigma}$ are defined
according to (16).
Then the $\alpha$-coefficients
$\alpha_{A_1;B_1;C_1|A_2;B_2;C_2}^{A_3;B_3;C_3}(n|N)$ are defined
as coefficients in front of the leading order $n$  terms in the OPE
\eqn\grav{\eqalign{
P^{n|N}_{A_1\phi+B_1\chi+C_1\sigma|A_2\phi+B_2\chi+C_2\sigma}(z)
e^{A_3\phi+B_3\chi+C_3\sigma}(w)\sim
\cr
{{\alpha_{A_1;B_1;C_1|A_2;B_2;C_2}^{A_3;B_3;C_3}(n|N)}\over{(z-w)^n}}
e^{A_3\phi+B_3\chi+C_3\sigma}(w)}}
(if the actual leading order of a given OPE is less than $n$, the
appropriate coefficient is zero)
Although the manifest form of the associate ghost polynomials (AGP)
is generally complicated, there is an algorithm
significantly simplifying the computations of both $AGP$ and
related $\alpha$-coefficients. 
The algorithm is based on comparison of two similar operator
products.
Below we shall explain the algorithm
and present the results for 
$\alpha_{3;1;0|1;-1;0}^{-3;0;1}(q-b|q)$ (where $0\leq{b}\leq{q}\leq{8}$)
and 
$\alpha_{2;-2;-1|0;1;0}^{-3;0;1}(n|8)$ entering the series (46).
Consider the operator product
\eqn\grav{\eqalign{
e^{A_1\phi+B_1\chi+C_1\sigma}(z)e^{A_2\phi+B_2\chi+C_2\sigma}(w)
=\sum_{m=0}^\infty
{{(z-w)^{-A_1A_2+B_1B_2+C_1C_2+m}}\over{m!}}
\cr\times
:e^{(A_1+A_2)\phi+(B_1+B_2)\chi+(C_1+C_2)\sigma}(z)
P^{(m)}_{A_1\phi+B_1\chi+C_1\sigma}:(w)}}
around the point $w$.
Differentiating $N$ times over $z$ we get
\eqn\grav{\eqalign{
\partial^{(N)}e^{A_1\phi+B_1\chi+C_1\sigma}(z)e^{A_2\phi+B_2\chi+C_2\sigma}(w)
\cr
=\sum_{m=0}^\infty\prod_{l=0}^{N-1}(-A_1A_2+B_1B_2+C_1C_2+m-l)
{{(z-w)^{-A_1A_2+B_1B_2+C_1C_2+m-N}}\over{m!}}
\cr\times
:e^{(A_1+A_2)\phi+(B_1+B_2)\chi+(C_1+C_2)\sigma}
P^{(m)}_{A_1\phi+B_1\chi+C_1\sigma}:(w)}}
On the other hand this product by definition coincides with the OPE:
\eqn\grav{\eqalign{
:P^{(N)}_{A_1\phi+B_1\chi+C_1\sigma}e^{A_1\phi+B_1\chi+C_1\sigma}:(z)
e^{A_2\phi+B_2\chi+C_2\sigma}(w)
\cr
=\sum_{j,k=0}^\infty\sum_{n=0}^N
{{(z-w)^{-A_1A_2+B_1B_2+C_1C_2+n-N+j+k}}\over{j!k!}}
\cr\times
:\partial^{(j)}
P^{n|N}_{A_1\phi+B_1\chi+C_1\sigma|A_2\phi+B_2\chi+C_2\sigma}
\cr\times
{P^{(k)}_{A_1\phi+B_1\chi+C_1\sigma}}e^{(A_1+A_2)\phi+(B_1+B_2)\chi+(C_1+C_2)\sigma}:(w)
}}
(the derivative of the associate polynomials appear since
in the definition (49) the polynomials are located at $z$
while the OPE (53) is around $w$).
This gives the characteristic equation on 
 $P^{n|N}_{A_1\phi+B_1\chi+C_1\sigma|A_2\phi+B_2\chi+C_2\sigma}$:
\eqn\grav{\eqalign{
\sum_{j,k=0}^\infty\sum_{n=0}^N
{{(z-w)^{-A_1A_2+B_1B_2+C_1C_2+n-N+j+k}}\over{j!k!}}
\cr\times
:\partial^{(j)}
P^{n|N}_{A_1\phi+B_1\chi+C_1\sigma|A_2\phi+B_2\chi+C_2\sigma}
{P^{(k)}_{A_1\phi+B_1\chi+C_1\sigma}}:
\cr
=
\sum_{m=0}^\infty\prod_{l=0}^{N-1}(-A_1A_2+B_1B_2+C_1C_2+m-l)
\cr\times
{{(z-w)^{-A_1A_2+B_1B_2+C_1C_2+m-N}}\over{m!}}
P^{(m)}_{A_1\phi+B_1\chi+C_1\sigma}}}

Matching the coefficients in front of each power of  $(z-w)$
(starting from the most singular term) 
then
gives recurrence relations  on
$:P^{n|N}_{A_1\phi+B_1\chi+C_1\sigma|A_2\phi+B_2\chi+C_2\sigma}:$
expressing them in terms of conformal
dimension $n$ combinations of
 $P^{(l)}_{A_1\phi+B_1\chi+C_1\sigma}$ and their derivatives
(with $1\leq{l}\leq{n}$).
The coefficients $\alpha_{A_1;B_1;C_1|A_2;B_2;C_2}^{A_3;B_3;C_3}(n|N)$
are then obtained by replacing
each of $\partial^{(k)}P^{(l)}$ according to
\eqn\grav{\eqalign{
P^{(l)}_{A_1\phi+B_1\chi+C_1\sigma}\rightarrow
(-1)^k
\prod_{i=0}^{k-1}\prod_{j=0}^{l-1}
(l+i)(-A_1A_3+B_1B_3+C_1C_3-j)}}
in each of these  combinations
since
\eqn\grav{\eqalign{
{P^{(l)}_{A_1\phi+B_1\chi+C_1\sigma}}(z)e^{A_3\phi+B_3\chi+C_3\sigma}(w)
\cr
\sim(z-w)^{-l}\prod_{j=0}^{l-1}
(-A_1A_3+B_1B_3+C_1C_3-j)e^{A_3\phi+B_3\chi+C_3\sigma}(w)+O(z-w)^{1-l}}}

\vfill\eject

Applied to $P^{n|8}_{2\phi-2\chi-\sigma|\chi}$ this procedure gives
the recurrence relations

\eqn\grav{\eqalign{P^{n|8}_{2\phi-2\chi-\sigma|\chi}=
9!P^{(n)}_{2\phi-2\chi-\sigma}\delta_{1;n}
-\sum_{k=1}^{n-1}\sum_{l=0}^k{{{\partial^{(l)}}
P^{k|8}_{2\phi-2\chi-\sigma|\chi}P^{(n-l-k)}_{2\phi-2\chi-\sigma}
}\over{l!(k-l)!}}
}}
and the corresponding $\alpha$-coefficients 
$\alpha_{2,-2,-1|0,1,0}^{-3,0,1}(n|8)$ are given by
\eqn\grav{\eqalign{
\alpha_{2,-2,-1|0,1,0}^{-3,0,1}(0|8)=9!
\cr
\alpha_{2,-2,-1|0,1,0}^{-3,0,1}(1|8)=-8!\times{40}
\cr
\alpha_{2,-2,-1|0,1,0}^{-3,0,1}(2|8)=8!\times{150}
\cr
\alpha_{2,-2,-1|0,1,0}^{-3,0,1}(3|8)=-8!\times{300}
\cr
\alpha_{2,-2,-1|0,1,0}^{-3,0,1}(4|8)=8!\times{275}
\cr
\alpha_{2,-2,-1|0,1,0}^{-3,0,1}(5|8)=-8!\times{94}
\cr
\alpha_{2,-2,-1|0,1,0}^{-3,0,1}(6|8)=0
\cr
\alpha_{2,-2,-1|0,1,0}^{-3,0,1}(7|8)=0
\cr
\alpha_{2,-2,-1|0,1,0}^{-3,0,1}(8|8)=0.}}
Similarly, when
applied to $P^{q-b|q}_{3\phi+\chi|\phi-\chi}$
$(0\leq{b}\leq{q}\leq{8})$ this procedure gives
the recurrence relations
\eqn\grav{\eqalign{
P^{q-b|q}_{3\phi+\chi|\phi-\chi}={{(q+3)!}\over{3!}}P^{(q)}_{3\phi+\chi}
(\delta_{1;q}+\delta_{2;q}+\delta_{3;q})
\cr
-\sum_{k=1}^{q-1}\sum_{l=0}^{k}
 {{{\partial^{(l)}}
P^{k|8}_{3\phi+\chi|\phi-\chi}P^{(n-l-k)}_{3\phi+\chi}
}\over{l!(k-l)!}}}}
and the corresponding $\alpha$-coefficients are
$\alpha_{3,-1,0|1,-1,0}^{-3,0,1}(q-b|q)(0\leq{b}\leq{q}\leq{8})$ are
\eqn\grav{\eqalign{
\alpha_{3,-1,0|1,-1,0}^{-3,0,1}(0|q)={{(q+3)!}\over{6}}\cr
(\alpha_{3,-1,0|1,-1,0}^{-3,0,1}(1|q)=-{3\over2}(q+2)!q)
\sum_{p=1}^8\delta_{p;q}\cr
\alpha_{3,-1,0|1,-1,0}^{-3,0,1}(2|q)=(-6(q+3)!+12(q+2)!q+72(q+1)!)
\sum_{p=2}^8\delta_{p;q}\cr
\alpha_{3,-1,0|1,-1,0}^{-3,0,1}(3|q)=(28(q+3)!-42(q+2)!q-504q!q)
\sum_{p=3}^8\delta_{p;q}\cr
\alpha_{3,-1,0|1,-1,0}^{-3,0,1}(4|q)
\cr
=(84(q+3)!
+126(q+2)!q-1512(q+1)!+3024q!q)
\sum_{p=4}^8\delta_{p;q}\cr
\alpha_{3,-1,0|1,-1,0}^{-3,0,1}(5|q)
\cr
=(-639(q+3)!-339(q+2)!q
+4896(q+1)!-7560q!q)
\sum_{p=5}^8\delta_{p;q}\cr
\alpha_{3,-1,0|1,-1,0}^{-3,0,1}(6|q)
\cr
=(1352(q+3)!+600(q+2)!q
-6984(q+1)!+10080q!q)
\sum_{p=6}^8\delta_{p;q}\cr
\alpha_{3,-1,0|1,-1,0}^{-3,0,1}(7|q)
\cr
=(1158(q+3)!-3918(q+2)!q
+7992(q+1)!-52920q!q)
\sum_{p=7}^8\delta_{p;q}\cr
\alpha_{3,-1,0|1,-1,0}^{-3,0,1}(8|8)=9!
}}
This fully determines the $\alpha$-coefficients
in the series (46).

\centerline{\bf Appendix B. BRST Relations and 
Gauge Transformations for 
$\omega^{2|1}$ and $\omega^{2|2}$.}

In this Appendix section we present explicit expressions
for BRST commutators leading to the gauge transformations
for the frame-like fields and relating frame-like and Fronsdal
fields for spin 3.
The gauge transformation for the $\omega^{2|1}$  field:
\eqn\lowen{
\omega_m^{ab|c}(p)\rightarrow\omega_m^{ab|c}(p)
+p_m\Lambda^{ab|c}(p)}
leads to shifting the $V^{2|1}$ vertex operator (21)
by BRST-exact terms:
$V^{2|1}(p)\rightarrow{V^{2|1}}(p)+\lbrace{Q},W_1^{2|1}(p)\rbrace$
where, up to overall numerical factor,
\eqn\grav{\eqalign{
W_1^{2|1}(p)\sim\Lambda^{ab|c}(p)\oint{dz}
ce^{-5\phi+ipX}((p\partial\psi)(\psi_c\partial^2{X_b}
-2\partial\psi_c\partial{X_b})+(p\psi)\partial^2\psi_c\partial{X_b})
\cr\times
({2\over{5}}\partial{L_a}\partial\xi-L_a\partial^2\xi)}}
where
\eqn\grav{\eqalign{L_a=\partial^2\psi_a-2\partial\psi_a\partial\phi
+{1\over{13}}\psi_a(5\partial^2\phi+9(\partial\phi)^2)}}
and $\Lambda$ has the same symmetry in the fiber indices as 
$\omega^{2|1}$.
This operator is BRST-exact if $\omega$ is transverse
in the $a,b$ fiber indices
(which, in turn, is the invariance condition).
Next, if $\omega_m^{ab|c}(p)$ is antisymmetric in $m$ and $a$
(so that the corresponding $\omega^{2|0}$ is the two-row field),
$V^{2|1}$ is again the BRST commutator in the small Hilbert
space:
\eqn\grav{\eqalign{
{V^{2|1}}(p)=\lbrace{Q},W_2^{2|1}(p)\rbrace}}
with
\eqn\grav{\eqalign{
W_2^{2|1}(p)\sim\omega_m^{ab|c}(p)\oint{dz}ce^{-5\phi+ipX}
(\psi_c\partial^2{X_b}-\partial\psi_c\partial{X_b})
\cr\times
({2\over{5}}\partial\psi^{\lbrack{m}}\partial{L_{a\rbrack}}
\partial\xi
-
\partial\psi^{\lbrack{m}}{L_{a\rbrack}}\partial^2\xi)
\cr
+\partial^2\psi_c\partial{X_b}
({2\over{5}}\psi^{\lbrack{m}}\partial{L_{a\rbrack}}
\partial\xi
-
\psi^{\lbrack{m}}{L_{a\rbrack}}\partial^2\xi)
}}
Next, we analyze $\omega^{2|2}$ and its vertex operator
(22).
The gauge transformation for the $\omega^{2|2}$  field:
\eqn\lowen{
\omega_m^{ab|cd}(p)\rightarrow\omega_m^{ab|cd}(p)
+p_m\Lambda^{ab|cd}(p)}
leads to shifting the $V^{2|2}$ vertex operator (21)
by BRST-exact terms:
$V^{2|2}(p)\rightarrow{V^{2|2}}(p)+\lbrace{Q},W_1^{2|2}(p)\rbrace$
with
\eqn\grav{\eqalign{
W_2^{2|2}(p)\sim\Lambda^{ab|cd}(p)\oint{dz}ce^{-6\phi+ipX}
\cr\times\lbrace
({1\over4}(p_n\partial{N^n})\partial\xi-
(p_nN^n)\partial^2\xi)
(\partial^2\psi_c\partial
^3\psi_{d}\partial{X^a}\partial{X_b}
-2\partial\psi_c\partial
^3\psi_{d}\partial{X_{a}}\partial^2{X_{b}}
\cr
+{5\over8}\partial\psi_c\partial
^2\psi_{d}\partial{X_{a}}\partial^3{X_{b}}
+{{57}\over{16}}
\partial\psi_c\partial
^2\psi_{d}\partial^2{X_{a}}\partial^2{X_{b}})
\rbrace
}}
where
\eqn\grav{\eqalign{
N_n=\partial^3{X_n}-{3\over2}\partial^2{X_n}
-{1\over3}\partial{X_n}((\partial\phi)^2-{{17}\over6}
\partial^2\phi)}}
As before, this operator is BRST-exact if $\omega$ is transverse
in the $a,b$ fiber indices.
Finally,
if $\omega_m^{ab|cd}(p)$ is antisymmetric in $m$ and $a$ or $b$
(so that the corresponding $\omega^{2|0}$ is the two-row field),
$V^{2|2}$ is again the BRST commutator in the small Hilbert
space:
\eqn\grav{\eqalign{
{V^{2|2}}(p)=\lbrace{Q},W_2^{2|2}(p)\rbrace}}
with
\eqn\grav{\eqalign{
W_2^{2|2}(p)\sim\omega_m^{ab|cd}(p)\oint{dz}ce^{-6\phi+ipX}
\lbrace
({1\over4}{N^m}\partial\xi-
(N^m)\partial^2\xi)
\cr\times
(\partial^2\psi_c\partial
^3\psi_{d}\partial{X^a}\partial{X_b}
-2\partial\psi_c\partial
^3\psi_{d}\partial{X_{a}}\partial^2{X_{b}}
\cr
+{5\over8}\partial\psi_c\partial
^2\psi_{d}\partial{X_{a}}\partial^3{X_{b}}
+{{57}\over{16}}
\partial\psi_c\partial
^2\psi_{d}\partial^2{X_{a}}\partial^2{X_{b}})
-(a\leftrightarrow{m})
\rbrace}}

\listrefs

\end